\documentclass[prx,twocolumn,showpacs,preprintnumbers,amsmath,amssymb,superscriptaddress]{revtex4}

\usepackage{graphicx}
\usepackage{dcolumn}
\usepackage{color}
\usepackage{multirow}

\newcommand{\beq}{\begin{equation}}
\newcommand{\eeq}{\end{equation}}
\newcommand{\beqa}{\begin{eqnarray}}
\newcommand{\eeqa}{\end{eqnarray}}

\begin{document}
\title{Band alignment and charge transfer in complex oxide interfaces}

\author{Zhicheng Zhong}
\affiliation{Max-Planck-Institut f\"ur Festk\"orperforschung,
  Heisenbergstrasse 1, 70569 Stuttgart, Germany}

\author{Philipp Hansmann}
\affiliation{Max-Planck-Institut f\"ur Festk\"orperforschung,
  Heisenbergstrasse 1, 70569 Stuttgart, Germany}


\begin{abstract}
The synthesis of transition metal heterostructures is currently one of the most vivid fields in the design of novel functional materials. 
In this paper we propose a simple scheme to predict \emph{band alignment }and \emph{charge transfer} in complex oxide interfaces.
For semiconductor heterostructures band alignment rules like the well known Anderson or Schottky-Mott rule are based on comparison of the work function or electron affinity of the bulk components. This scheme breaks down for oxides due to the invalidity of a single workfunction approximation as recently shown (Phys. Rev. B 93, 235116; Adv. Funct. Mater. 26, 5471). Here we propose a new scheme which is built on a continuity condition of valence states originating in the compounds' shared network of oxygen. It allows for the prediction of sign and relative amplitude of the intrinsic charge transfer, taking as input only information about the bulk properties of the components. We support our claims by numerical density functional theory simulations as well as (where available) experimental evidence. Specific applications include i) controlled doping of SrTiO$_3$ layers with the use of 4$d$ and 5$d$ transition metal oxides and ii) the control of magnetic ordering in manganites through tuned charge transfer.
\end{abstract}
\date{\today}
\maketitle

\section{Introduction}
Until today semiconductors present the most important class of functional
materials for electronic applications. Their usage in electronic components
and the continuous development of new devices keeps on pushing the limits of technology. 
In almost all such devices the key functionality originates not in the physics of the bulk, but in the peculiarities of
interfaces \cite{Kroemer:rmp01}. Yet, semiconductor devices have intrinsic limitations: i) the characteristic length 
scales are relatively large so that further downscaling (current state of the art is 7$nm$ technology) becomes very unlikely 
and Moore's law is bound to end; ii) solely charge degrees of freedom are exploited. Transition metal oxides (TMO) on the other hand 
provide spin, orbital, charge and lattice degrees of freedom \cite{Imada:rmp98,Tokura:sc00} and are therefore viewed as one of the best candidates to replace semiconductors in future electronic device. Thanks to an immense progress of epitaxial growth techniques, TM oxides heterostructures can now be controlled on atomic length scales. 
Several novel physical phenomena have been discovered in recent years and potential multi-functional devices seem to be realizable\cite{Mannhart:sci10, Takagi:sc10, Triscone:review2011, Hwang:natm12, Chakhalian:Natm12}.

In oxide electronics, one of the cornerstone mechanisms in complex heterostructures (in analogy to semiconductors) is the alignment of bands at a hetero interface and, driven by the resulting potential gradient, a charge transfer across the interface\cite{Hwang:Nat02,Ohtomo:nat04,Dagotto:JPCM08,
  Logvenov:sc09,May:Natm09,Barriocanal:NatCom10,
  Barriocanal:adv10,Bhattacharya:prl10,Gibert:Natm12,Chen:prl13,
  Kleibeuker:prl14, Salafranca:prl14, Chen:nl15, Grisolia:Natp16,
  Cao:NatCom16,Wang:sc15,Nichols:NC16}. This mechanism can be seen as intrinsic doping without undesired disorder induced by chemical doping. Consequently, controlling 
heterostructures with a wide variety and range of experimentally tunable parameters (e.g. strain, thickness, substrate choice, etc.) allows to 
engineer new phases which do not exist in bulk. It is, hence, obvious that predictive power for the direction and amplitude of 
charge transfer with rules that are as simple as possible is highly desirable. 

\begin{figure}[t]
  \includegraphics[width=0.5\textwidth]{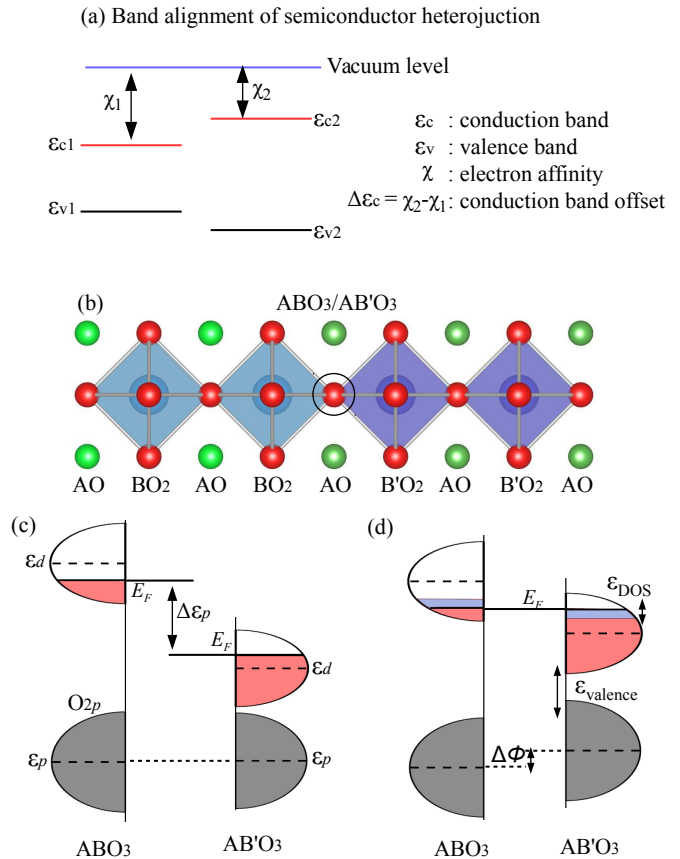}
  \caption{\label{Fig1} 
(a)Schematic figure of Anderson's rule in semiconductor heterojunction: the vacuum energy levels of the two semiconductors on either side are aligned. (b) Schematic figure of a perovskite (001) interface ABO$_3$/AB$^\prime$O$_3$
emphasizing the common network of oxygen sites across the interface. (c)
Alignment of oxygen states at the interface would generally yield mismatch in
the heterostructure's Fermi energy (see Eq.~\ref{eq1}). This mismatch drives a
charge transfer which itself alters interface and local potentials. The final
equilibrium state is shown in (d) with indications (black arrows) of the three
different contributions in the energy balance equation
Eq.~\ref{eq2}. $\varepsilon_p$ and $\varepsilon_d$ are the local energy levels
of oxygen $p$ and TM $d$ states.}
\end{figure}

A natural first attempt would be the usage of well established semiconductor rules, such as Anderson's or the Schottky-Mott rule \cite{Davies:book}. As schematically shown in Fig. \ref{Fig1}(a) the vacuum energy levels of the two semiconductors on either side of the heterojunction should be aligned, which rely on the electron affinity (or work function) of semiconductors. However, those rules are not suitable for TM oxides, due to the fact that the reduction of the work function to a single value is an approximation that does not hold for TM oxides \cite{zhong:prb16, Jacobs:AFM16}. Moreover, the characteristic lengthscales in oxides are one or two orders of magnitude smaller than in semiconductors, and hence certain approximations are no longer justified so that non-trivial microscopic terms need to be taken into account explicitly.

In this paper we propose a rule that is based on the continuity of states in the TMO's oxygen matrix and which allows for qualitative prediction of band alignment and charge transfer in complex TMO heterostructures. The oxygen continuity boundary
condition allows us to explain and predict the induced charge transfer between
the constituents of the heterostructure starting only from energetics of the
bulk compounds.  In the first part of our report we sketch the underlying
driving forces in TMO hetero compounds which are built from perovskites ABO$_3$
with A being the a cation (e.g. Sr or La) and B as a TM from strongly
correlated 3$d$, 4$d$, or strongly spin-orbit coupled 5$d$ shell. We claim that
bulk data for the oxygen 2$p$ energies $\varepsilon_p$ of the components can
be used for predictions of ABO$_3$/AB$^\prime$O$_3$ interfaces, and prove it
by showing quantitative data for a wide range of materials. In the second part
of the paper we provide selected examples as a proof of principle, and
predictions for possible devices yet to be synthesized.

\section{Methods}
To reveal general material trends, we first study bulk complex transition
metal oxides ABO$_3$ with a cubic perovskite structure (in the majority of cases
A is taken as Sr but we also discuss other cases like Ca and La) with B being
a 3$d$ (Ti-Co), 4$d$(Zr-Rh) or 5$d$ (Hf-Ir) transition metal element. The
lattice constant is fixed at the optimized value of SrTiO$_3$,
$a$=3.945\AA \footnote{The DFT-GGA optimized lattice constant for SrTiO$_3$ is $a$=3.945\AA, which is about 1$\%$ larger than the experimental one. Our estimates of effects from strain show that such effects are negligible for the conclusions we draw.}.  We then use SrRuO$_3$ as an example for effects of strain,
cation A substitution, structural distortion, and magnetism. Most
calculations were carried out for superlattices of
(ABO$_3$)$_n$/(AB$^\prime$O$_3$)$_n$ with n=5 and n=1, in order to estimate
possible quantum confinement effects for the latter. Moreover, in order to validate the
assumption of clean interfaces we performed simulations of a rough interface with a 25$\%$
cation mixture (for details see appendix). It turns out that the observed changes are
indeed small and will not affect our conclusions. For the study of
magnetism in SrMnO$_3$ heterostructures, we take into account an on-site
Coulomb interaction U$_{Mn}$=2eV and a realistic GdFeO$_3$-type structural
distortion which are considered to be important for a realistic description of
the magnetism. In all calculations the atomic positions are fully relaxed. 

Density functional theory (DFT) calculations were performed with the
VASP (Vienna ab initio simulation package) code \cite{Kresse:prb99} using the
generalized gradient approximation GGA -PBE functional \citep{PerdewPRL96} for
electronic exchange and correlation. We consider that energy separation
between transition metal $d$ states and oxygen 2$p$ states is usually
underestimated in the PBE potential, and use the MBJ potential
\cite{Tran:prl09}, which is implemented in wien2k \cite{WIEN2k}. We perform
Wannier projection \cite{Kune20101888, Mostofi2008685} of the oxygen $p$ Bloch
states and TM $t_{2g}$ $d$ states to obtain an accurate value of the local
energy levels $\varepsilon_p$ and $\varepsilon_d$ \cite{Haverkort:prb12} as
shown in Fig.~\ref{Fig4} and Table ~\ref{Tableone}. Be aware that $\varepsilon_p$ and $\varepsilon_d$ indicate the local energy terms of $t_{2g}$ and oxygen $p$ states respectively and not to the center of gravity for valence or conduction bands ($\varepsilon_v$ and $\varepsilon_c$) \cite{Noguera:book}. They coincide only in the ionic limit, but for finite covalence (i.e. oxygen-TM hybridization) $\varepsilon_v$ and $\varepsilon_c$ are split apart and the bands have no pure oxygen or TM character anymore. The values in Fig.~\ref{Fig4} correspond to $\varepsilon_d$ and $\varepsilon_p$. It turns out that covalence yields an energy drop of approximately 0.4eV of $\varepsilon_p$ for all materials listed in the Fig.~\ref{Fig4}. We have also carefully checked that the projection details such as including $e_g$ orbitals will not change our conclusion.

In order to obtain quantitative estimates for the charge transfer we integrate the electron density below the Fermi level (down to the energy gap between $d$ and oxygen $p$ states) projected inside atomic spheres. The radius of the atomic spheres (and here is the mentioned ambiguity in the definition) were chosen to the default values of PAW potential in VASP, e.g. 1.2 Angstrom for a Vanadium atom in SrVO$_3$.  In this way, the integrated electron density of Vanadium atom in bulk SrVO$_3$ is 0.75. Considering the formal valence of vanadium to be one d electron, we then take 0.75 as renormalizing factor which we also apply to the interface case in order to obtain a quantitative value for the charge transfer. Moreover, in order to estimate the sensitivity of such values with respect to the choice of the radius of atomic spheres, we have checked that a 10$\%$ increase of the radius only induces approximately a 3$\%$ change of the charge transfer estimated in this way. Hence, the numbers for the charge transfer given in Tab.~\ref{Tableone} have to be taken with some care. We stress that our conclusions do not rely on specific quantitative values, but on overall trends.

\section{Intrinsic charge transfer in complex oxide interfaces}

Let us start the discussion with the idea that motivated the study. For the
sake of simplicity we restrict ourselves to perovskite ABO$_3$
heterostructures along the (001) direction. As common in such TM oxide
compounds the most relevant states are i) the empty or partially filled TM $d$
states at the Fermi energy \footnote{In band insulating materials such as SrTiO$_3$, it turns out (from several DFT simulations of the material) that the assumption of a slight $n$-type doping ($E_F$ at the bottom of the conduction band) is the appropriate choice. } which are split into $t_{2g}$ and $e_g$ states in
cubic perovskites and ii) the oxygen 2$p$ states residing at an energy
$\varepsilon_p$ a few eV below the Fermi energy and forming more or less
covalent bonds with the TM $d$ states. 
At an interface of two materials with
different lattice constants or distortion patterns of an ideal cubic case we
typically find a smooth transition where such structural features mutually
propagate between the components of the heterostructure \cite{Rondinelli:am11,
  Liao:Natm16}. A rather natural observation is that such smooth structural
transitions in the shared oxygen lattice necessarily demand a continuity of
the oxygen states across the interface layers as a boundary
condition.

\begin{figure}[t]
  \includegraphics[width=0.5\textwidth]{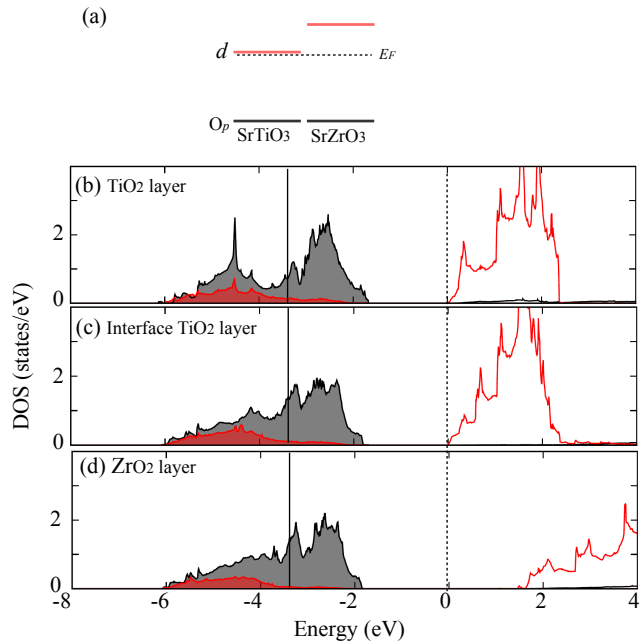}
  \caption{\label{Fig2} 
Sketch of the energy level scheme (a) and layer resolved DFT density of states
for the (SrTiO$_3$)$_{5}$/(SrZrO$_3$)$_{5}$ superlattice (b,c,d). Oxygen $p$
states (dark gray) and TM $d$ states (red) are shown with the Fermi level
marked by the dotted black line. Additionally we show as a solid black line
the layer dependent center of mass of the oxygen states which corresponds to
the layer dependent $\varepsilon_p^i$. We plot the DOS for the two BO$_2$
layers directly at the interface as well as for the TiO$_2$ two unit cell
further away. The alignment of oxygen states in this band insulator
heterostructure is practically perfect.}
\end{figure}

\begin{figure}[t]
  \includegraphics[width=0.5\textwidth]{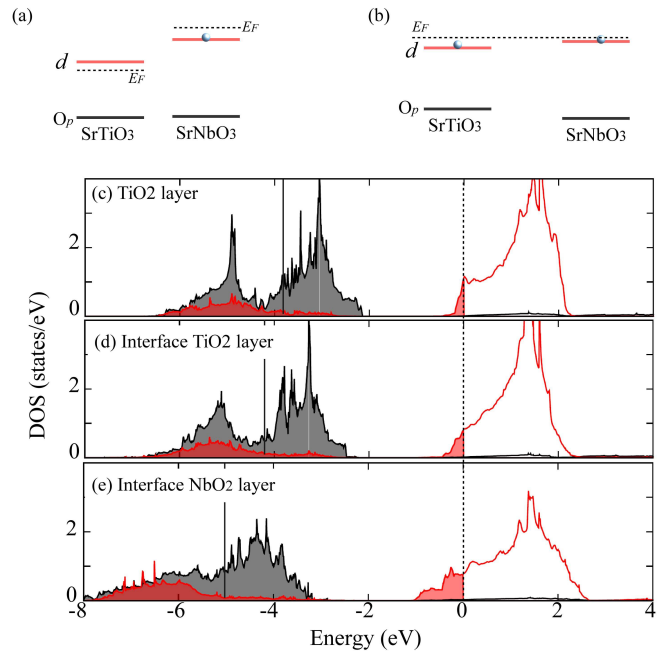}
  \caption{\label{Fig3} 
Upper panel: Sketch of the energy level scheme before (a) and after (b) electron
transfer; Lower panels: Same plots as in Fig.~\ref{Fig2} now for the
heterostructure (SrTiO$_3$)$_{5}$/(SrNbO$_3$)$_{5}$. Opposed to the
(SrTiO$_3$)$_{5}$/(SrZrO$_3$)$_{5}$ we see clear electron transfer across the
interface and the expected structure in the alignment of the layer dependent
$\varepsilon_p^i$ in the shown layers. The transferred electron into interfacial and bulk TiO2 layers are 0.23$e$ and 0.17$e$ respectively.} 
\end{figure}

One might start by visualizing the oxygen continuity condition with a very
simple sketch we show in Fig.~\ref{Fig1}(b) for an ABO$_3$/AB$^\prime$O$_3$
interface; later on we will extend our discussion to more general cases with
different cations. In a hypothetical two step procedure the continuity
condition would demand that at the interface the oxygen states need to be
lined up which, for two materials with different $\varepsilon_p$, would result
in a  mismatch of the Fermi energy $E_F$ equivalent to
\begin{equation}
\label{eq1}
\Delta \varepsilon_p=\varepsilon_{p}^{\rm { ABO_{3}}}-\varepsilon_p^{\rm {AB^\prime O_{3}}}
\end{equation}
, see Fig.~\ref{Fig1}(c). Since $E_F$ must, however, be constant all through
the heterostructure in equilibrium, a charge transfer occurs between the
layers which itself creates i) an electrostatic potential drop $\Delta\phi$
across the interface, ii) rigid band shifts we indicate by
$\Delta\varepsilon_{\rm DOS}$, and iii) a local electrostatic potential drop
$\Delta\varepsilon_{dp}$ yielding relative shifts between TM $d$
and oxygen $p$. These three terms counter the original driving potential - see
Fig.~\ref{Fig1}(d). As indicated in this last sketch we end up with a balance
of potentials that has to be calculated self consistently. Before further
formalization, let us support this hypothesis by numerical calculations
starting with the easiest case of interfacing two band insulators which, in
fact, has a one to one correspondence to the band alignment mechanism in
semiconductor heterostructures.  

In Fig.~\ref{Fig2} we show the layer dependent density of states for the
interface of SrTiO$_3$ and SrZrO$_3$ simulated in a 5/5 superlattice. The
chemical potential, which is basically free to move inside the gap, is
indicated by the dashed black line while the center of mass (i.e. the onsite
energy) of the oxygen 2$p$ states (dark gray DOS) is shown as a black solid
line. In the plots, we show ZrO$_2$ and TiO$_2$ layers directly at the
interface as well as the TiO$_2$ bulk-like layer two unit cells further away
from the interface. As can be seen, the alignment of oxygen states across the
interface and even in the next TiO$_2$ layer is practically perfect due to the
absence of any charge transfer between the insulating layers. This observation
is a first indication of the validity of the backbone hypothesis of the oxygen
states continuity. We note in passing that we have an advantage compared to
semiconductor heterostructures where obvious structural and electronic
continuity is absent\cite{Tersoff:prb84}. In such semiconducting
heterojunctions one might try to employ the  Anderson's (or Schottky-Mott) rules in order to estimate the band mismatch by the difference of the
components electron affinity (or work function) \cite{Davies:book}. It turns out, however, that
in realistic cases this rules often fails to predict band offsets. Furthermore, a
direct extrapolation of this rule to oxides is rather questionable due to the
fact that the concept of a compounds work function is invalid as we have shown in a recent study of
oxide heterostructure workfunctions\cite{zhong:prb16, Jacobs:AFM16} which
strongly depend on details of the surface orientation and termination. In
contrast, the oxygen state continuity rule is a much stronger boundary
condition and, as we point out in several occasions, its predictions seem to
agree well with experimental observation.

Now we take one step further and consider SrNbO$_3$ instead of SrZrO$_3$ in
the same geometry, i.e. a metal with one valence electron in the Nb 4$d$
shell\cite{Oka:prb15}, which we show in Fig.~\ref{Fig3} - here we see that an
initial alignment of oxygen states has lead to electron transfer from Nb 4$d$-
to Ti 3$d$-states across the interface. The self consistently determined final
position of oxygen states with respect to the Fermi level is in 
agreement with the expectations from our sketch in Fig.~\ref{Fig1}(d).
I.e., the transferred electron has lead to an additional potential $\Delta\phi$
in the interface region and changes in the local potentials (discussed in more
detail below) which eventually leads to a balanced monotonous evolution of the
layer dependent energy of oxygen 2$p$ states. Let us stress, that in our
picture the directions and relative strength of the charge transfer seem to be
predictable just by comparison of $\Delta \varepsilon_p$ of the bulk
ABO$_3$. 

Before providing numerical data and materials calculations to confirm
this, let us give more formal arguments. Indeed, the concept of band alignments
at interfaces is well known and understood in semiconducting $pn$-junctions
\cite{Ashcroft_and_Mermin}, where one observes charge modulations in the so
called ''space charge'' region of the typical order of $\approx100$~\AA. In
the $pn$-junctions two identical semiconductors are interfaced which, however
have a Fermi level mismatch due to different, either p- or n-type, kind of
doping that is resolved by charge transfer. Yet, in our case there are some
crucial differences specific to oxide interfaces: First of all our
length-scales are at least one order of magnitude smaller and the variation of the
induced potential on the order of a few \AA ~ prohibits clearly a
semiclassical model commonly used for semiconductors. Secondly, we have to
consider more microscopic details for the energy balance equation which
finally determines $\Delta n_e$. Namely, in addition to $\Delta\phi$ (the only
term of relevance for semiconductor $pn$-junctions) the charge transfer in
oxides induces shifts in the local potentials of the different TM sites which
might be disentangled into contributions (i) from mutual change
of the valence of B and B$^\prime$ site $\Delta\varepsilon_{dp}$
yielding relative shifts between TM $d$- and oxygen 2$p$- states with a sign
equal to that of $\Delta n_e$ and ii) from the specific structure of density
of states $\Delta\varepsilon_{\rm DOS}$. In summary we can write
\begin{equation}
\label{eq2}
-\Delta \varepsilon_p = \Delta\phi+\Delta\varepsilon_{\rm DOS}+\Delta\varepsilon_{dp}
\end{equation}
The terms on the right hand side depend on the transferred charge $\Delta n_e$
and we may try to linearize them. For the first term $\Delta\phi$ we can
assume a plate capacitor model which would yield a potential drop per unit
cell of $\Delta\phi=\Delta n_e \cdot d/\epsilon$ with $d$ being the effective
distance of charge transfer cross the interface and $\epsilon$ the dielectric
permitivity \citep{Nakagawa:Natm06,Zhong:prb10}. The second term can be
simplified by assuming an approximately constant density of states around
Fermi level $E_F$ to be $\Delta n_e \cdot D$, where
$D=\frac{1}{D_B(E_F)}+\frac{1}{D_{B^\prime}(E_F)}$,  $D_B$ and $D_{B^\prime }$
the local density of states for B and B$^\prime$ sites. This contribution D
can be seen from Fig.~\ref{Fig3} to be of the order of up to $1$eV. The last term indicates the change of $\varepsilon_{dp}$ induced by the charge transfer that modifies the valence of transition metal.The argument for the linearity in $\Delta n_e$ of the last term is easily
understood by considering a Hartree type self energy ($\propto n_e$) so we can
assume $\Delta\varepsilon_{dp}\approx \Delta n_e \cdot U_H$. $U_H$ reflects the change of the energy due to the static single particle mean-field energy that comes from electronic Coulomb interaction. Using virtual crystal approximation for SrVO$_3$ allows us to roughly estimate this
contribution to be also of the order of $1$eV. Hence, assuming the charge
transfer of the order of $\Delta n_e$ =1 will lead, different to
semiconductors, to non-negligible contributions of D and $U_H$, since the
first term $d/\epsilon$, i.e. the typical length scale, is much smaller in our
oxide heterostructures than in semiconductors. We arrive eventually at a
simplified linear relation between $\Delta n_e$ and $\varepsilon^{\rm bulk}_p$:
\begin{equation}
\label{eq3}
\Delta n_e \approx -\frac{1}{(d/\epsilon + D + U_ H)}\cdot\Delta \varepsilon_p
\end{equation}
While also the simplified equation is hardly solvable in a closed form it
allows for a remarkable insight and confirmation of
our initial idea: The sign and strength of the charge transfer at an interface
between two materials, which turns out to be much larger than in semiconductor
devices, should be determined by the difference of the respective bulk oxygen
2$p$ energies with respect to their Fermi level $\varepsilon_p$.\footnote{Let
  us, finally, remark that this relation also holds for the case of band
  insulating interfaces (i.e. vanishing density of states at $E_F$) where D
  will diverge and, hence an infinitesimal $\Delta n_e$ leads to finite
  $\Delta\varepsilon_p$}.

\begin{figure}[t]
  \includegraphics[width=0.5\textwidth]{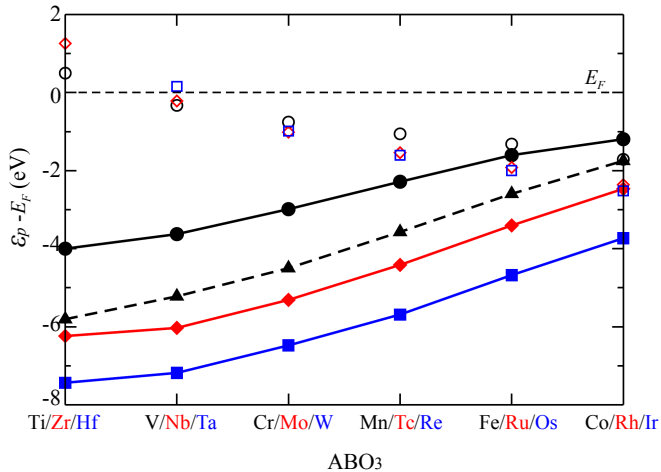}
  \caption{\label{Fig4} 
Summary of bulk $\varepsilon_{p}$ (filled symbols) and $\varepsilon_{d}$
(empty symbols) with respect to the Fermi level ($E_F$=0) for different
SrBO$_3$ (solid line) materials. For the B site we consider 3$d$ (black), 4$d$
(red), and 5$d$ (blue) elements. With this reference bulk data the charge
transfer in heterostructures consisting of these materials can be
anticipated. The simple criterion for the direction of the charge transfer at
the ABO$_3$/AB$^\prime$O$_3$ interface is that the component with lower
$\varepsilon_{p}$ will donate electrons to the other one.  Additionally we
plot data for LaBO$_3$ (dashed line) for B=3$d$ which is used for estimates
for ABO$_3$/A$^\prime$BO$_3$ interfaces. (for numerical data of the plot and
additional compounds please refer to Tab.~\ref{Tablethree} in the appendix)}
\end{figure}

With this insight we generate DFT reference data for a variety of bulk
compounds. The results are summarized in Fig.~\ref{Fig4} where we plot the
average energy of oxygen 2$p$ states (filled symbols) and the average energy
of partially filled d-orbitals (empty symbols) with respect to the Fermi
level \footnote{The change of $E_F$ within the series is quite small, e.g $E_F$(SrVO$_3$)=5.20eV,  $E_F$(SrMnO$_3$)=5.13eV} for SrBO$_3$ (solid lines) with B being a 3$d$ (black), 4$d$ (red), or 5$d$ (blue) element. For the 3$d$ series we additionally show oxygen $p$
energies for LaBO$_3$ (dashed line). The figure nicely shows clear trends in
the series of materials considered:
\begin{itemize}
\item Within the intra 3$d$, 4$d$, and 5$d$ series we observe a monotonous and
  almost linear increase of $\varepsilon_p$ of about $2-3$eV within each period.\\
\item Within one group we have a monotonous drop in $\varepsilon_p$ for
  a given configuration of about $\geq 3$ eV from 3$d$ to 5$d$ compounds.\\
\item Changing Sr to La, i.e. decreasing the oxidation number of the
  transition metal leads to a decrease of $\varepsilon_p$ up to two eV. This
  change is actually closely related to $\Delta\varepsilon_{\rm valence}$ in
  Eq.~\ref{eq2} and represents in some way the extreme case where we have
  changed the nominal charge by 1. We note, moreover, that a change of the
  cation with identical oxidation state ($\{$Sr, Ca, Ba$\}$ or $\{$ La, Y$\}$)
  leaves the results for $\varepsilon_p$ basically unchanged, see table
  Tab.~\ref{Tablefour} in the appendix.\\
\end{itemize}

Before we turn to the discussion of how these numbers can be used for
predictions we need to address the fact that the results shown in
Fig.~\ref{Fig4} are all obtained for undistorted cubic unit cells. The energy
scale of the listed material trends is of the order of a few eV and it should
be emphasized that certain effects beyond idealized structures will lead to
modifications of $\varepsilon_p$ on energy scales that are non-negligible
compared to our reference data. Therefore we have performed
calculations for specific cases studying the influence of strain, orthorhombic
distortions, and magnetism (in Tab.~\ref{Tablefour} in the appendix we provide
the numerical data for these benchmark cases). 

Starting with strain, we see for the example of SrRuO$_3$, that 1$\%$
compressive strain will decrease the $\varepsilon_p$ of clearly non-negligible
0.2eV and realize that this effect can actually be additionally exploited to
tune the energetics for the desired effect \citep{Chen:prb14,
  Tseng:jap16}. Next, we turn to orthorhombic distortions
\citep{Grisolia:Natp16} in the same material for which we also observe a
change of $\varepsilon_p$ of the order of $0.3$eV compared to the cubic case
which is less but still important for a reliable prediction. Finally, let us
address the influence of magnetic order for ferromagnetic SrRuO$_3$. Here we
see a split between up and down states not only in the Ru 4d states but also
in the associated oxygen states of $0.7$eV. We remark that this issue does not
occur in antiferromagnetic ordered structures and also for temperatures higher
than the Curie temperature. In the FM ordered phase a prediction of the charge
transfer by our simplified scheme is not straight forward and becomes
questionable if the corresponding energy scales are equal or larger than the
$\Delta\varepsilon_{p}$ in question.

After these remarks, however, we will now show that for many cases the
numerical data for the bulk materials shown in Fig.~\ref{Fig4} can be
exploited in order to predict the charge
transfer in layered heterostructures composed of the listed
materials. Starting with the compounds shown in Figures~\ref{Fig2} and
\ref{Fig3}, i.e. SrTiO$_3$ interfaces with Sr$\{$Zr, Nb$\}$O$_3$ we start with
the bulk values $\varepsilon_{p}^{\rm { SrTiO_{3}}}$=-4.0eV,
$\varepsilon_{p}^{\rm { SrZrO_{3}}}$=-6.23eV, and $\varepsilon_{p}^{\rm
  {SrNbO_{3}}}$=-6.03eV. For the SrTiO$_3$/SrZrO$_3$ interface both materials
show empty TM $d$ states and an alignment of $\varepsilon_{p}^{\rm
  {SrTiO_{3}}}$ with the lower lying  $\varepsilon_{p}^{\rm { SrZrO_{3}}}$
does not result in any electron transfer after equalizing the Fermi levels in
line with Eq.~\ref{eq3} due to the diverging $D$ contribution in the right
hand side denominator. For the second case, however, we observe that for
SrNbO$_3$ the oxygen states are much lower in energy than those of SrTiO$_3$,
i.e. $\Delta \varepsilon_p=2.03$eV $\gg 0$. Hence, an alignment of oxygen
states would lead to a mismatch of the Fermi energy such that electron needs to
be transferred from Nb 4$d$-states to the Ti 3$d$-states for a constant
equilibrium chemical potential across the interface. We can exploit the above
reported trends in order to make general predictions:\\

\begin{itemize}
\item Given the monotonous trends within early materials of the same period in
  the SrBO$_3$ series we conclude that electron transfer will be only possible
  from lighter to heavier B compounds. This is somewhat counter intuitive: For
  instance, $d^1$ SrVO$_3$ will transfer its electron to $d^2$  SrCrO$_3$  but
  not to the empty Ti $3d$ states of SrTiO$_3$. As it turns out, this is in
  agreement with experiment and other numerical simulations for a variety of
  interfaces including LaMnO$_3$/LaNiO$_3$ \citep{Gibert:Natm12,
    Dong:prb13,Lee:prb13,Hoffman:prb13},
  LaTiO$_3$/LaFeO$_3$ \cite{Kleibeuker:prl14,Zhang:prb15,He:prb16},
  LaTiO$_3$/LaNiO$_3$ \cite{Chen:prl13,Cao:NatCom16}, and SrVO$_3$/SrMnO$_3$
  \cite{Chen:prb14,Tseng:jap16}.\\
\item Given the monotonous behaviour of $\varepsilon_p$ within one group over
  different periods we can state that also here electron will be transferred
  from the heavier to the lighter B element. e.g. SrNbO$_3$ will electron dope
  SrVO$_3$. \\
\item For the more general case of ABO$_3$/AB$^\prime$O$_3$ one has to
  consider precise values to make predictions. For instance, SrIrO$_3$ should
  be able to intrinsically dope SrMnO$_3$ at an interface, while SrRuO$_3$
  cannot dope SrTiO$_3$.\\
\end{itemize}

\begin{figure}[t]
  \includegraphics[width=0.5\textwidth]{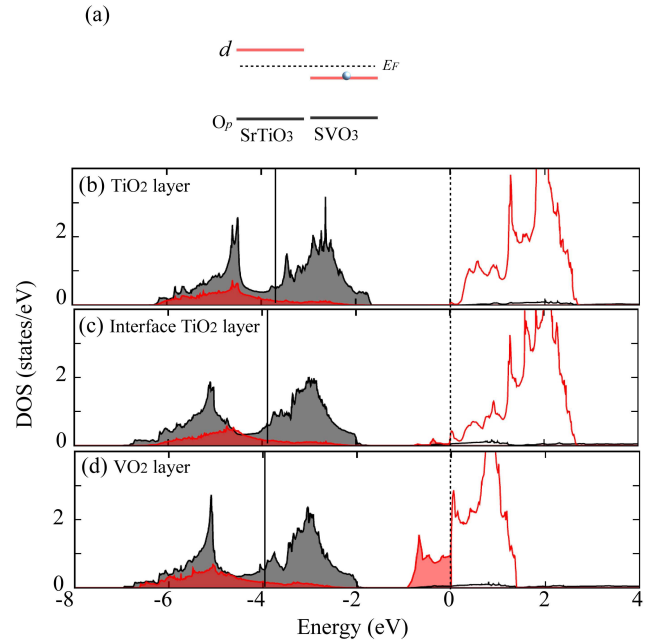}
  \caption{\label{Fig5} 
Same plot as in Figs.~\ref{Fig2} and \ref{Fig3} for a
(SrTiO$_3$)$_5$/(SrVO$_3$)$_5$heterostructure. In this case alignment of
oxygen bands will not drive electron transfer and (as expected from
Fig.~\ref{Fig4}) SrVO$_3$ is a bad candidate for doping SrTiO$_3$}
\end{figure}

In order to strengthen the first claim of this list we extend our simulations
with a calculation of a SrTiO$_3$/SrVO$_3$ interface for which we have a
$\Delta \varepsilon_p=-0.37$eV. Indeed, consistent with our arguments we see
from Fig.~\ref{Fig5} that practically no charge transfer occurs. The layer
resolved partial density of states shows that basically all $d$ electron remains in
the SrVO$_3$ slab and the oxygen 2$p$ states are rather well
aligned. \footnote{At a closer look one should note a very small amount of
  charge at the interface TiO$_2$ (concomitant with a non-perfect oxygen
  alignment) layer which one might attribute to a   covalent hybridization of
  Ti and V d-states directly at the interface. However, as we will see in the
  next step, this effect is very small compared to the charge transfer one can
  obtain by replacing SrVO$_3$ with another compound.}

At this point it might be asked if our analysis can be applied also to
periodic stacks which are further away from the limit of a single interface
than the so far considered 5/5 stacks. In order to answer this question we
performed several simulations of (SrBO$_3$)$_1$/(SrB$^\prime$O$_3$)$_1$
heterostructures and summarize the results in
Tab.~\ref{Tableone}. Remarkably, we find that in \emph{all} cases the charge
transfer can been anticipated in direction and relative amplitude with the
data from Fig.~\ref{Fig4}. This is remarkable since it is not a prior clear
that a qualitative argument made for the interface between two bulk materials
still holds for a periodic stack with strong quantum confinement effects
\cite{zhong:prb13} and lattice relaxations at the interfaces. They are fully
taken into account in the heterostructure calculations but do not spoil the
predictive capabilities from bulk calculations neither. Looking closer at the
values in Tab.~\ref{Tableone} reveals a remarkable consistency between $\Delta
\varepsilon_p$ and $\Delta n_{e}$. We have already seen in the example of the
SrTiO$_3$/SrVO$_3$ interface that in order to dope electron into Ti or V $3d$
states, $4d$ or $5d$ TM components have to be considered since a positive
$\Delta \varepsilon_p$ is needed and, as we can see from the stacks in the
first four rows $\Delta n_{e}$ grows monotonously with $\Delta
\varepsilon_p$. The same is true for the opposite case (reported in the lower
four rows) where an increased negative $\Delta \varepsilon_p$ drives an
increased transfer from the vanadium $3d$ states to heavier elements like Cr,
Mn, Fe, and Co in the $3d$ series. Moreover, we can learn from the observation
$\Delta \varepsilon_p > \varepsilon^{i}_{p}({\rm B})-\varepsilon^{i}_{p}({\rm
  B}^\prime)$, that for the energy balance (Eq.~\ref{eq2}) besides $\Delta
\phi$, $\Delta\varepsilon_{\rm DOS}$ and $\Delta\varepsilon_{\rm valence}$
contribute significantly.

\begin{table}[t]
\begin{ruledtabular}
\caption{Sign and trends of electron transfer in
  (SrBO$_3$)$_1$/(SrB$^\prime$O$_3$)$_1$ heterostructures. We list the
  energies of oxygen 2$p$ states resolved by BO$_2$
  ($\varepsilon^{i}_{p}$(BO$_2$)), AO ($\varepsilon^{i}_{p}$ (AO)) and B$^\prime $O$_2$
  layers ($\varepsilon^{i}_{p}$(B$^\prime$O$_2$)) with respect to the Fermi energy
  as well as the induced electron transfer $\Delta n_e$ from B to B$^\prime$
  sites. $\Delta \varepsilon_p$ taken from Fig. ~\ref{Fig4} is also listed.}
\label{Tableone}
\begin{tabular}{lrrrrrrr}
B/B$^\prime$   & $\varepsilon^{i}_{p}$(BO$_2$) & $\varepsilon^{i}_{p}$(AO) & $\varepsilon^{i}_{p}$(B$^\prime $O$_2$)&   $\Delta n_{e}$ & $\Delta \varepsilon_p$ \\
                         \hline
Ti/Nb & -4.67  & -4.93  & -5.22 & 0.36 & 2.03 \\
Ti/Ta & -5.07 & -5.46 & -5.89 & 0.63 &  3.18\\
V/Nb & -4.26 & -4.50 & -4.89 & 0.49 & 2.38 \\
V/Ta & -4.62 & -4.92 & -	5.62 & 0.71 & 3.56 \\
V/Cr & -3.72 & -3.61 & -3.60 &  -0.27 & -0.64 \\
V/Mn & -3.40 & -3.18 & -3.07 & -0.30 & -1.34 \\
V/Fe & -3.02 & -2.77 & -2.31 &  -0.49 & -2.02 \\
V/Co & -2.65 & -2.42 & -1.38 & -0.57   & -2.43
\end{tabular}
\end{ruledtabular}
\end{table}

Before the end of this section where we have focused mostly on interfaces of
the form SrBO$_3$/SrB$^\prime$O$_3$ let us make some remarks about the
generalization to ABO$_3$/A$^\prime{\rm BO}_3$ and the most general
ABO$_3$/A$^\prime {\rm B}^\prime$O$_3$ interfaces. In Fig.~\ref{Fig4} we have
already shown data for LaBO$_3$ in comparison to the SrBO$_3$ series. If we
consider the same element on the B site our scheme is directly applicable as
before. More specifically, $\varepsilon_{p}$ of a LaBO$_3$ perovskite is
always lower than in the corresponding SrBO$_3$ so that at a LaBO$_3$/SrBO$_3$
interface the electron is always transferred from LaBO$_3$ to SrBO$_3$ which is
consistent with experiments, e.g. for the case of LaTiO$_3$/SrTiO$_3$
\cite{Hwang:Nat02,Okamoto:Nat04} or LaMnO$_3$/SrMnO$_3$
\cite{May:Natm09,Bhattacharya:prl10}.

Additional complications arise in the most general ABO$_3$/A$^\prime$B$^\prime
$O$_3$ setup where we have two qualitatively different interface
configurations which might be either AO-B$^\prime$O$_2$ or
BO$_2$-A$^\prime$O.  We might actually consider the realistic case of a
SrTiO$_3$/LaMnO$_3$ interface \citep{Barriocanal:NatCom10,Barriocanal:adv10,
  Chen:Natm15, Wang:sc15}. Starting with the interface TiO$_2$-LaO we assume
to have a unit of LaTiO$_3$ linked with SrTiO$_3$ on the one side and
LaMnO$_3$ on the other - so we have broken the problem down to a
SrTiO$_3$/LaTiO$_3$ and a LaTiO$_3$/LaMnO$_3$ interface which we know how to
handle. As a result, by considering our reference data, we would arrive at a
prediction of charge transfer at the interface from the Ti states of the
central LaTiO$_3$ part to its neighbors. The opposite is true for the other
possible interface SrO-MnO$_2$, and in this case a decomposition as before
would yield the conclusion that no charge transfer will occur. This line of
argument is in agreement with experiments on SrTiO$_3$/LaMnO$_3$ where electron
transferred has been observed for the LaO-TiO$_2$ terminated interface
\cite{Barriocanal:adv10}. According to our claim, the other interface
structure, SrO-MnO$_2$, which experimentally can be realized by inserting a
monolayer of SrMnO$_3$, should not allow for such charge transfer. Please note we did not consider the contribution from polar discontinuity \cite{Noguera:jpcm00, Nakagawa:Natm06}, which might induce complexity such as oxygen vacancies and other defects.

We conclude our first main section with the claim that charge transfer at the
interfaces of perovskite heterostructures can be qualitatively predicted by
comparison of bulk qualities. It is not even necessary to perform a numerical
simulation of the interface in order to anticipate the direction and relative
magnitudes of charge transfer.

\section{Applications and predictions}
In the second part of our paper we highlight two applications of our
prediction scheme starting with the controlled doping of SrTiO$_3$ (see
\ref{dSTO}) drawing also a parallel to heterogeneous semiconductor devices. We
then turn to magnetic transitions in manganite heterostructures triggered by
controlled charge transfer (see \ref{magMn}) explaining past theoretical and
experimental results in the context of our unifying concept which we then
complement by predictions on how to improve external control of such
transitions.

\subsection{Electron doping in SrTiO$_3$ heterostructure}
\label{dSTO}

\begin{figure}[t]
  \includegraphics[width=0.5\textwidth]{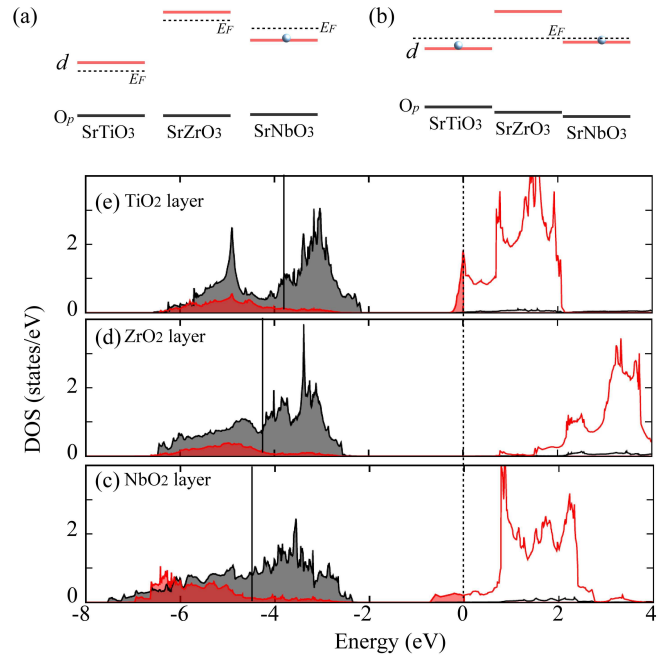}
  \caption{\label{Fig6} 
Same plot as in Fig.~\ref{Fig3} but now including an additional buffer layer
between SrTiO$_3$ and SrNbO$_3$. The unit cell of this three-component
symmetric heterostructure has the form
(SrTiO$_3$)$_{5}$/(SrZrO$_3$)$_{2}$/(SrNbO$_3$)$_{1}$/(SrZrO$_3$)$_{2}$. As
can be seen the electron transfer from Nb to Ti remains intact even after
inclusion of the buffer layer excluding a pivotal role of microscopic details
in the orbital degrees of freedom at the SrTiO$_3$/SrNbO$_3$ interface.}
\end{figure}

As a first application we chose to highlight how to control doping of one of
the most widely spread and used TM oxide compounds SrTiO$_3$. So far SrTiO$_3$ has
been doped by various 3$d$ TM oxides e.g. LaAlO$_3$ \citep{Ohtomo:nat04} and
LaTiO$_3$\citep{Hwang:Nat02} to produce a two dimensional electron gas but
there is no known SrBO$_3$ with B from the $3d$ series. With the data from
Fig.~\ref{Fig4} we have already given the reason why this is not surprising
and, in fact, impossible. The very naive guess that electron from a more filled
$3d$ element like, e.g. vanadium  would spill into the empty Ti $3d$ states is
prohibited by the energy balance that originates in the oxygen states
continuity condition (see Fig.~\ref{Fig5}). So, SrVO$_3$ is identified as a
bad candidate for doping SrTiO$_3$. Instead we have seen in Fig.~\ref{Fig3}
that SrNbO$_3$ is a much more promising component for that purpose. In
(SrTiO$_3$)$_5$/(SrNbO$_3$)$_5$ the continuity of oxygen 2$p$ states drives a
sizable charge transfer. What we have not discussed yet, however, is the
question if microscopic details of orbital degrees of freedom at the
interface, which are not included in our scheme, are equally or even more
important for the charge transfer than the considered energetics. In order to
shed light on this issue we extend our calculations with the inclusion of an
insulating buffer layer which will allow for a spatial separation of Ti and Nb
$d$-states. The choice of a perovskite buffer layer is actually quite straight
forward in the light of our reference data and we chose SrZrO$_3$ for this
purpose which will remain band insulating in the three component
heterostructure. In Fig.~\ref{Fig5} we show the result for a symmetric
(SrTiO$_3$)$_{5}$/(SrZrO$_3$)$_{2}$/(SrNbO$_3$)$_{1}$/(SrZrO$_3$)$_{2}$
superlattice which is experimentally realizable\cite{Beck:apl00,
  Moreira:CPL09, Kajdos:apl13,Chen:nl15}.
  
Here we see another confirmation of our simple argument: while the
buffer layer remains band insulating we still observe a transfer of electron
from Nb to Ti $d$-states even though they are spatially separated. At this point we might
refer to an analogous technique in semiconductor setups where the doping
impurities are spatially separated from the charge carriers. A concept which is commonly
known as ``modulation doping'' and widely applied in so called MODFET setups
\cite{Dingle:APL78, Takashi:jjap80}. Besides the shown results we have checked
the dependence of the charge transfer amplitude as a function of the thickness
of the buffer layer but were unable to find significant trends up to 
supercells that include buffers of 5 bulk unit cells. In fact modulation doped oxides
recently started to attract some theoretical \cite{Zhong:prb10} and
experimental \cite{Kajdos:apl13, Chen:Natm15,Chen:nl15} researchers.

\subsection{AFM to FM transition in SrMnO$_3$ heterostructures}
\label{magMn}
After having exemplified how our oxygen energetics argument can be used to
anticipate charge transfer in the case of SrTiO$_3$ based heterostructures we
now take one step further by considering manganite heterostructures. Here the
goal is similar yet more ambitious since a provoked charge transfer in
SrMnO$_3$ setups should have a sensitive impact on magnetic
transitions. Starting point for us is bulk SrMnO$_3$ which is found to have a
G-type antiferromagnetic ground state. Our goal is to find an appropriate
heterostructure in which doping into Mn $d$-states triggers a transition to a
ferromagnetic ground state - see review papers \cite{Coey:ap99,
  Dagotto:Physicsreport01, Salamon:rmp01, Tokura:rpp06} and references
therein. From our reference data we see that it should actually be rather easy
to provoke electron transfer into Mn $d$-states considering the comparatively
large value of $\varepsilon_p$ for Mn. Indeed it turns out that by employing
our predictive scheme we can confirm experimentally known routes to
charge-transfer induced magnetic transition and classify possible
heterostructure components to tune more or less charge transfer.

We carried out calculations for 1/1 and 2/2 layered superlattice
(SrMnO$_3$)$_{1}$/(SrBO$_3$)$_{1}$ (considering also a GdFeO$_3$ structural
distortion since magnetism can be strongly coupled to the lattice) for four
different species of B site elements. The results are summarized in
Tab.~\ref{Tabletwo} where we report the size of the magnetic moment on the B
and the Mn site of the respective compound. We have sorted the table  with ascending electron
transfer. As anticipated choosing SrIrO$_3$ as interfacing partner will result
in a little electron transfer and, in the 2/2 structure the antiferromagnetic
order with a moment of 3.05 $\mu_B$ remains. In the case of the 1/1 structure
we observe a somewhat larger electron transfer which is already sufficient to
trigger a magnetic transition to a ferromagnetic ground state
\citep{Nichols:NC16} with ordered Mn moments of 3.24 $\mu_B$. The next
component we consider is SrVO$_3$ - here the data from Fig.~\ref{Fig2} tells
us to expect similar to the Ir compound only a small amount of transferred
electron which we see confirmed in our numerical results. Interestingly, both 1/1
and 2/2 compounds seem to be intermediate between antiferromagnetic and
ferromagnetic phase yielding a ferrimagnetic ground
state\cite{Chen:prb14,Tseng:jap16}. We can further predict which components would
yield a much  more significant electron transfer: Candidates are easily
nominated from Fig.~\ref{Fig4} and to this end we select SrNbO$_3$, which is
recently grown experimentally\cite{Oka:prb15},  as well as SrTaO$_3$ - our
most potent electron donor for oxide heterostructures. The selected components
perform as predicted in the numerical simulations and we observe a large
charge transfer pushing the (SrMnO$_3$)$_{1,2}$/(SrNbO$_3$)$_{1,2}$ and
(SrMnO$_3$)$_{1,2}$/(SrTaO$_3$)$_{1,2}$ into a ferromagnetic metallic ground
state with ordered Mn moments up to 3.74$\mu_B$.\footnote{Let us remark once
  more that, due to additional confinement effects, the 1/1 compounds are the
  materials ``furthest away'' from a single interface situation so that some
  exceptions to the trends are quite expected - it is actually rather
  remarkable how well most of the predictions from bulk energetics still
  work!}

\begin{table}[t]
\begin{ruledtabular}
\caption{Magnetic ground states of (SrBO$_3$)$_n$/(SrMnO$_3$)$_n$ with
  $n$=1,2. B = Ir, V, Nb, and Ta. The magnetic moments localized on TM B and
  Mn sites are also listed.}
\label{Tabletwo}
\begin{tabular}{lllllll}
       & \multicolumn{3}{l}{1/1} & \multicolumn{3}{l}{2/2} \\
       &  B ($\mu_B$)     & Mn($\mu_B$)    & M      & B($\mu_B$)      & Mn($\mu_B$)    & M  \\
                          \hline
SrIrO$_3$ & 0.0    & 3.24  & Ferro  & 0.0    & 3.05  & AFM   \\
SrVO$_3$  & -0.19  & 3.2   & Ferri  & -0.36  & 3.16  & Ferri  \\
SrNbO$_3$ & 0.05   & 3.66  & Ferro  & 0.02   & 3.56  & Ferro  \\
SrTaO$_3$ & 0.03   & 3.74  & Ferro  & 0.0    & 3.67  & Ferro  
\end{tabular}
\end{ruledtabular}
\end{table}

\begin{figure}[t]
  \includegraphics[width=0.5\textwidth]{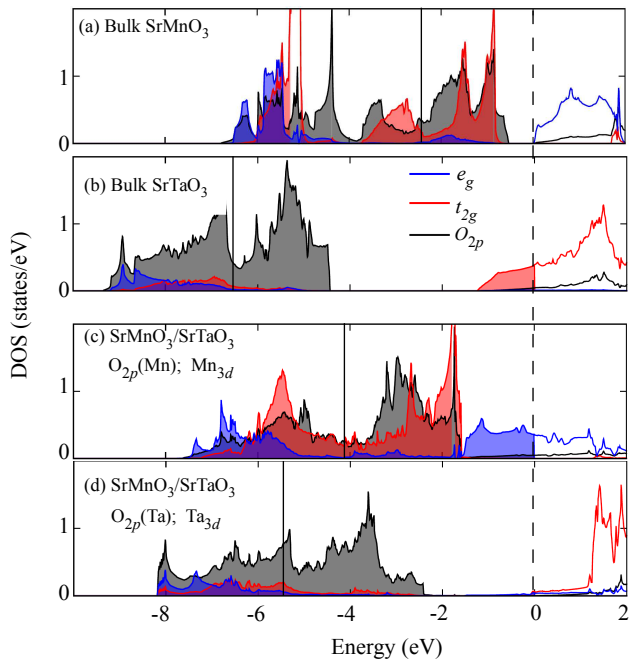}
  \caption{\label{Fig7} 
Plot of the  density of states with the usual color/marker convention
(including now also 3$d$-$e_g$ states plotted in blue). We show results for
magnetically ordered calculations starting with bulk results in the upper two
panels. SrMnO$_3$ is G-type antiferromagnetic ordered and SrTaO$_3$
paramagnetic. In the lower two panels we plot the data for the two layers in a
(SrMnO$_3$)$_{1}$/(SrTaO$_3$)$_{1}$ heterostructure including a GdFeO$_3$
distortion. We see the dramatic influence of the electron transfer from Ta to Mn
and the induction of ferromagnetic ordered metallic MnO$_2$ layers.}
\end{figure}

In order to underline the data given in Tab.~\ref{Tabletwo} we plot the
resolved single particle density of states in Fig.~\ref{Fig7} - this time,
however, for magnetic DFT calculations. In the upper panels of the figure we
start as a reference with the bulk results for the gapped antiferromagnetic
DOS  of SrMnO$_3$ and metallic and paramagnetic SrTaO$_3$. In the lower two
panels one can nicely observe how in the (SrMnO$_3$)$_{1}$/(SrTaO$_3$)$_{1}$
heterostructure almost all Ta $d$ electron is depleted in the MnO$_2$ layers
yielding a metallic ferromagnetic ground state with significant filling of
the Mn 3$d$-$e_g$ states.

\section{Summary and Outlook} 
In summary we have presented a simple prediction scheme for the tuning of
charge transfer in oxide heterostructures. By employing arguments based
on the requirement that the $p$ states of the oxygen network in a
heterostructure need to be continuous, we were able to utilize reference data
from bulk calculations in Fig.~\ref{Fig4} to predict the direction and
relative amplitude of charge transfer between layers in oxide
heterostructures. Remarkably this scheme remains in tact even for layered
compounds down to a 1/1 geometry. As a proof of principle we provided
simulations for electron doping of SrTiO$_3$ in heterogeneous arrays
simultaneously confirming experimentally known trends and offering suggestions
to material growers which are the most promising components to control the low
energy electronic structure. In a second application part we have exemplified, with the help of SrMnO$_3$ based
heterostructures, how controlled charge transfer can be systematically
exploited in order to trigger magnetic phase transitions. The
shown examples are only a taste of what can be predicted with the unifying
concepts we presented and we hope that experimental colleagues will be
inspired by our work to create new materials and devices. This includes
specifically also the possibility to couple/entangle - via charge transfer -
physics of $3d$ and $5d$ TM oxides \cite{Nichols:NC16,Yin:PRL13} which,
each on their own, are governed by very different energy scales. One might
try, for instance, to induce spin-orbit coupling effects in $3d$ systems,
e.g. in the case of SrMnO$_3$ /SrTaO$_3$ where one might find a large magnetic
anisotropy energy and antisymmetric exchange on the magnetic Mn sites. One
might even target the realization of yet more exotic skyrmion, spin-spiral,
and topological phases. On the computational side we already started a hunt
for the best possible candidates to be manipulated towards exciting new ground
states.

In order to give an outlook for future theory and computational development we
state once more that for certain cases the application of our scheme is not
straight forward and needs to be extended. One of these cases is the
interfacing of compounds that are ferromagnetic ordered in bulk. Another issue
that should be mentioned is the change of $\varepsilon_p$ due to correlation
effects: Given that the relevant effects we discussed have a characteristic
energy scale of a few eV, shifts due to single particle self energies might
have to be included instead of the simplified $\Delta\varepsilon_{dp}$ contribution in Eq.~\ref{eq2}. A very clear example where this
will be the case is the usage of charge transfer insulators, typically found in late 3$d$
TM oxides e.g. cuprates. An extension to include such
correlation effects is very desirable since the sensitivity of correlated
electron systems would add even more possibilities to generate novel
functionality like, e.g. the recently suggested Mott
transistor\cite{Zhong:prl15}. Moreover, focusing more on material trends the
presented initial study did not extensively discuss how to use the thickness
of periodic stacks or buffer layers to quantitatively tune charge transfer and
how to interface non-perovskite oxides. The oxygen continuity
condition, or adaptations thereof, should be valid not only in the complex
oxides with perovskite structures, but also in other lattice structures like
anatase, pyrochlores, spinel, Ruddlesden-Popper and double perovskites
\cite{Tomioka:prb00, Kato:prl03, Long:Nat09} which are structurally equal to
the 1/1 interfaces along a (111) direction.
\section{Acknowledgments}
We thank H. Boschker, J. Nichols, H. N. Lee, Y. Lu, L. Si, Z. Liao, Z. Wang, Y. Cao
and J. Mannhart for motivation and valuable discussions.

\appendix

\section{Simulation of a rough surface}
\begin{figure}[t]
  \includegraphics[width=0.5\textwidth]{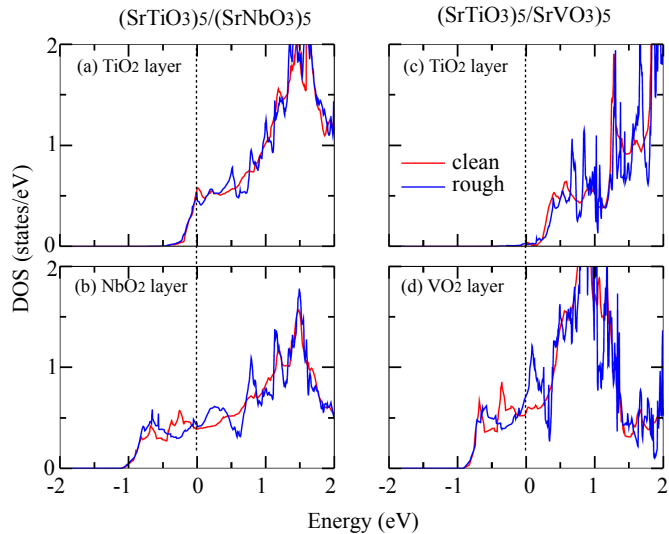}
  \caption{\label{Fig8}
Projected Density of TM $t_{2g}$ states of  the BO$_{2}$ layer with two layers away from interface. Red indicates a clean interface, while blue indicates a rough interface with 25$\%$ cation intermixing.}
\end{figure}

To test the effect of cation mixing we have performed benchmark calculations with up to 25$\%$ mixing.
To this end we employ a 2$\times$2 supercell with a layered structure as -BO$_2$-AO-B$_{0.75}$B$^\prime_{0.25}$O$_2$-AO-B$_{0.25}$B$^\prime_{0.75}$O$_2$-AO-B$^\prime$O$_2$. The comparison between the density of states for clean and rough interfaces with (and without) charge transfer is shown in Fig.\ref{Fig8}. Our main conclusion remains, hence, unchanged, and effects like cation mixing can indeed be regarded as a secondary effect.

\section{Numerical data}
In Tab.~\ref{Tablethree} we provide the numerical data plotted in
Fig.~\ref{Fig4}. Moreover, in Tab.~\ref{Tablefour} we report values of
$\varepsilon_{p}$ for benchmark calculations including effects of cation
exchange, orthorhombic distortion, strain, and ferromagnetic ordering.

\begin{table}[H]
\begin{ruledtabular}
\caption{ Numerical data of Fig.~\ref{Fig4} $\varepsilon_{p}$ of SrBO$_3$ with B=3$d$, 4$d$, and 5$d$. The unit is eV}
\label{Tablethree}
\begin{tabular}{llllll}
3$d$   & $\varepsilon_{p}$ & 4$d$  & $\varepsilon_{p}$ &  5$d$ & $\varepsilon_{p}$ \\
                         \hline
SrTiO$_3$ &  -4.00 & SrZrO$_3$  &  -6.23 & SrHfO$_3$  &  -7.43                    \\ 
SrVO$_3$ & -3.63  & SrNbO$_3$  & -6.03 & SrTaO$_3$  &   -7.18                   \\                         
SrCrO$_3$ & -2.99   & SrMoO$_3$  & -5.31  & SrWO$_3$  & -6.48                     \\ 
SrMnO$_3$ & -2.28  & SrTcO$_3$  & -4.41 & SrReO$_3$  &   -5.69                   \\ 
SrFeO$_3$ & -1.60  & SrRuO$_3$  & -3.40 & SrOsO$_3$  &    -4.68                  \\ 
SrCoO$_3$ & -1.19  & SrRhO$_3$  & -2.47  & SrIrO$_3$  &     -3.73      
\end{tabular}
\end{ruledtabular}
\end{table}

\begin{table}[H]
\begin{ruledtabular}
\caption{Reference data for $\varepsilon_{p}$ of manganites and ruthenates
  depending on cation exchange, othorhombic distortion, compressive strain and
  ferromagnetic ordering. The unit is eV}
\label{Tablefour}
\begin{tabular}{llll}
Manganites   & $\varepsilon_{p}$ & Ruthenates  & $\varepsilon_{p}$  \\
      \hline
KMnO$_3$ & -1.47    &KRuO$_3$ &  -2.48 \\
CaMnO$_3$ & -2.30    &CaRuO$_3$ &  -3.44\\
BaMnO$_3$ & -2.22    &BaRuO$_3$ &  -3.29\\
LaMnO$_3$ & -3.57    &LaRuO$_3$ &  -4.56\\
SrMnO$_3$ & -2.28    &SrRuO$_3$ &  -3.40 \\
1$\%$ compressive SrMnO$_3$ & -2.42  & SrRuO$_3$ & -3.60 \\
distorted CaMnO$_3$ & -2.70   & SrRuO$_3$   & -3.72 \\
FM SrMnO$_3$ &  -2.25/-2.75 & SrRuO$_3$  &  -3.46/-4.17
\end{tabular}
\end{ruledtabular}
\end{table}


\end{document}